\documentclass{aa}
\usepackage{graphicx}
\begin{document}
%

\title{Optical Spectroscopy of IRAS 02091+6333\thanks{Based on
observations at the Innsbruck 60cm telescope and  the Flemish
Mercator telescope on Roque de los Muchachos, Spain}}

\author{S. Kimeswenger\inst{1} \and A. Bacher\inst{1} \and M. Emprechtinger\inst{1}
\and G.E. Gr{\"o}mer\inst{1} \and  W. Kapferer\inst{1} \and  W.
Kausch\inst{1} \and \protect\newline M.G. Kitzbichler\inst{1} \and
M.F.M. Lechner\inst{1} \and C. Lederle\inst{1} \and K.
Uytterhoeven\inst{2,3} \and  A.A. Zijlstra\inst{4} \phantom{XX}}

\offprints{S. Kimeswenger,\\
~~ \email{stefan.kimeswenger@uibk.ac.at}}

\institute{Institut f\"ur Astrophysik der
Leopold--Franzens--Universit\"at Innsbruck, Technikerstr. 25, 6020
Innsbruck, Austria \and Instituut voor Sterrenkunde, Katholieke
Universiteit Leuven, Celestijnenlaan 200 B, 3001 Leuven, Belgium
\and Mercator Telescope, Calle Alvarez de Abreu 70, E-38700 Santa
Cruz de La Palma, Spain \and Astrophysics Group, Department of
Physics, UMIST, P.O. Box 88, Manchester M60 1QD, UK}

\date{Received  22 April 2003/ Accepted 16 June 2003}

\abstract{We present a detailed spectroscopic investigation,
spanning four winters, of the asymptotic giant branch (AGB) star
IRAS~02091+6333.  Zijlstra \& Weinberger (\cite{ZW}) found a giant
wall of dust around this star and modelled this unique phenomenon.
However their work suffered from the quality of the optical
investigations of the central object. Our spectroscopic
investigation allowed us to define the spectral type and the
interstellar foreground extinction more precisely. Accurate multi
band photometry was carried out. This provides us with the
possibility to derive the physical parameters of the system. The
measurements presented here suggest a weak irregular photometric
variability of the target, while there is no evidence of a
spectroscopic variability over the last four years.

 }

\maketitle
%

\section{Introduction}
When Zijlstra \& Weinberger (\cite{ZW}) discovered a massive dust
shell around the AGB star IRAS~02091+6333 they could use only a
single quick look spectrum and the original TYCHO  $B_{\rm T}$ and
$V_{\rm T}$ magnitudes (Perryman et al. \cite{tycho}) to estimate
the spectral type and the interstellar extinction towards the
target. The spectroscopic investigation was lacking comparison
spectra obtained with the same instrument setup. Furthermore the
TYCHO instrument was close to its sensitivity limits. Thus we
obtained spectra and photometry of this unique object for several
years to derive an accurate spectral type and the foreground
extinction. This allowed us to determine more precisely the
distance of the target for the modelling of the dust shell found
on Infrared Astronomical Satellite (IRAS) images. Zijlstra \&
Weinberger (\cite{ZW}) outline such shells for various types of
objects at late stages of their evolution. Their focus was
especially on a swept up shell with a void in the interstellar
matter (ISM) around the target. This is crucial for both, the
"Swiss cheese" like structure of the ISM and for the hydrodynamic
evolution of the planetary nebula (PN) built after this
evolutionary stage.

We thus investigated the spectroscopic and photometric properties
of the central star to provide new input for the complete model of
this unique object and its surroundings.

\section{Data}
The spectroscopic data were obtained with the Innsbruck 60~cm
telescope (Kimeswenger \cite{60cm}; Bacher et al. \cite{bacher01})
and an OptoMechanics 10C spectrograph. A CompuScope Kodak 0400 CCD
camera was attached to the spectrograph. In the year 2000 a
grating with 600~l/mm was used, resulting in a resolution of
1\,\AA/pixel. Later a grating with 240~l/mm (2.6\,\AA/pixel) was
mounted to achieve a better S/N ratio. The complete log of the
observations is given in Table~\ref{tab_obs}.
 The exposure times
varied with the wavelength region from 300 to 900 seconds. Each
spectrum is a composite of individual takes covering about
800\,\AA\ with the high resolution grating and 2000\,\AA\ with the
medium resolution mode. The wavelength shifts were selected to
obtain appropriate overlaps of at least 25\% of the wavelength
range. In most of the nights several spectra were obtained to
improve the S/N and eliminate cosmic ray events. The positions
along the slit and the starting wavelengths varied to eliminate
possible systematic errors in the setup. Flatfield, bias
subtraction and wavelength calibration were carried out in a
standard manner with the help of MIDAS routines. For comparison
the late type standard stars  HD~23475 (M2IIb), HD~39801 (M2I),
HD~13325(M3III), HD~40239 (M3II), HD~44478 (M3III), HD~42995
(M3III) HD~5316 (M4III) and HD~12292 (M5III) were observed with
the medium resolution setup. The spectra were not calibrated
absolutely, but only corrected for exposure time and airmass (with
the atmospheric extinction given by T\"ug \cite{extinct}).

The wide band direct imaging in Innsbruck was obtained with an
AP7p SITe 502e CCD device (Kimeswenger et al. \cite{v838}; Lederle
\& Kimeswenger \cite{CI}). 24 images were taken in the nights of
10$^{\rm th}$ and 11$^{\rm th}$ of December 2002 with $B$, $V$,
$R$ and $I_{\rm C}$ filters. The exposure times were 30, 20, 10
and 10 seconds in $B$, $V$, $R$ and $I_{\rm C}$ respectively.
After standard basic CCD reduction the source extraction was
performed using SExtractor V2 (Bertin \& Arnouts \cite{sex}). The
rms of the comparison standards in the field was $< 0\fm01$
($\sigma_B=0\fm003$, $\sigma_V=0\fm009$, $\sigma_R=0\fm007$ and
$\sigma_I=0\fm004$). Absolute calibration was obtained in both
nights using the standard stars HR~580 and HR~596. The rms
variations of the zero-points were less than 0\fm03, resulting in
an error of the mean zero-point of less than 0\fm015.

\begin{table}[ht]
\caption{Log of the Innsbruck 60cm observations} \label{tab_obs}

\begin{tabular}{ccccc}
\hline
Date & mode & band / & resolution \\
& & wavelength & sampling & eff.$^*$ \\
& & & \AA/pixel & \AA \\
 \hline \hline
05.02.2000 & spectra & 6000-9000\AA & 1.0 & 6.2\\
15.01.2001 & spectra & 5000-10000\AA & 2.6 & 10.6 \\
18.01.2001 & spectra & 5000-10000\AA & 2.6 &  10.6 \\
11.02.2001 & spectra & 5440-10000\AA & 2.6 & 9.9 \\
15.02.2001 & spectra & 5000-8860\AA & 2.6 & 8.6 \\
16.02.2001 & spectra & 5440-8860\AA & 2.6 & 8.0 \\
06.03.2001 & spectra & 5440-8860\AA & 2.6 & 10.7 \\
08.01.2002 & spectra & 5600-10000\AA & 2.6 & 13.4 \\
09.01.2002 & spectra & 5000-9000\AA & 2.6  & 14.1\\
10.12.2002 & imaging & BVRI$_{\rm C}$ &  \\
11.12.2002 & imaging & BVRI$_{\rm C}$ &  \\
17.01.2003 & spectra & 5000-10000\AA & 2.6 & 17.1 \\
\hline
\end{tabular}\\
$^*$ {\scriptsize Effective resolution was measured as FWHM of
night sky lines.}
\end{table}
The long term monitoring of the target was done with the P7
photometer attached to the 1.2m Mercator telescope located at La
Palma, using the 7 filters of the  Geneva photometric system
(Meylan \& Hauck \cite{mercator}, Golay \cite{golay}). The data
was obtained and reduced within the framework of monitoring stars
with constant airmass as described in Burki et al. (\cite{burki}).
The typical errors derived by the rms of the standard stars for
the nights are below 0\fm01. The results are listed in
Table~\ref{mercator_tab}.
\begin{figure}[ht]
\centering
\includegraphics[width=\columnwidth]{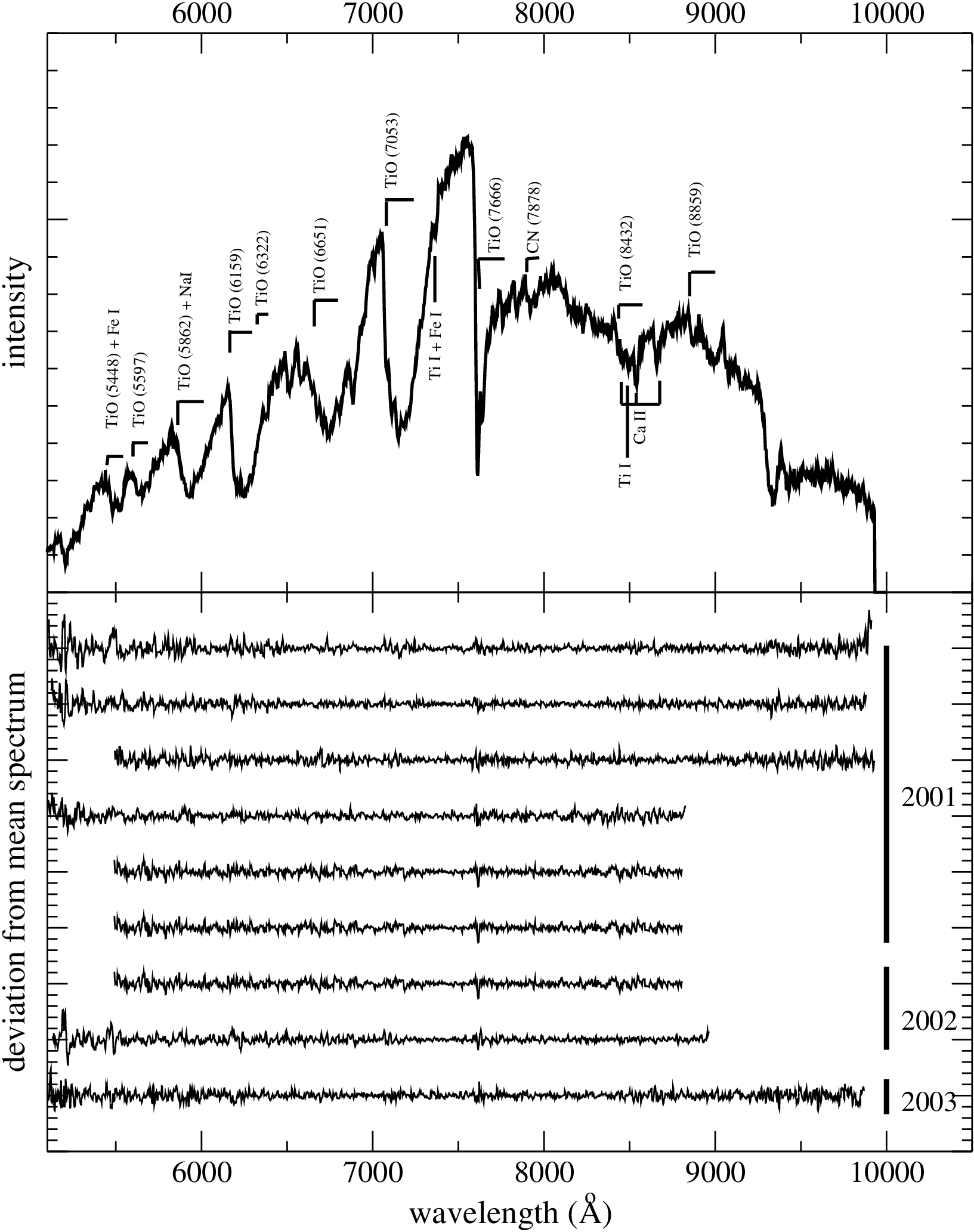}
\caption{The average spectrum (upper panel) with the most dominant
features identified and the deviation from the mean spectrum of
those spectra, using the same instrument setup, during winter
2001, 2002 and 2003. For clarity an offset of 1.0 was added to
each spectrum. No spectroscopic variability in the lines or in the
continuum trends can be found.} \label{fig_all}
\end{figure}


\section{Results}

\subsection{Spectroscopy}

First the spectra were investigated with respect to variability.
We cannot detect spectral variability (except in the region of the
telluric water vapor bands longwards to 8000\AA -- see
Fig.~\ref{fig_all}). Also individual equivalent widths of metal
lines of the IR \ion{Ca}{II} triplet and of \ion{Mg}{} lead us to
the same conclusion. Thus a composite spectrum was derived for the
classification. The classification criteria of Kirkpatrick et al.
(\cite{KP}) were applied in the same way as described in
Kimeswenger et al. (\cite{iras}). This results in a spectral class
of M4 - M5. The scheme of TiO indices (Malyuto et al. \cite{SK1})
argues for an M4.2III star. Applying the extension by
Schmidt-Kaler \& Oestreicher (\cite{SK2}) we obtain an M4III star
with  $M_{bol} \approx -2\fm65$ ($M_{V} \approx -0\fm5$). The
relation of the TiO to the NaD and Mgb indices propose luminosity
class III.

\begin{figure*}[ht]
\sidecaption
 \includegraphics[width=108mm]{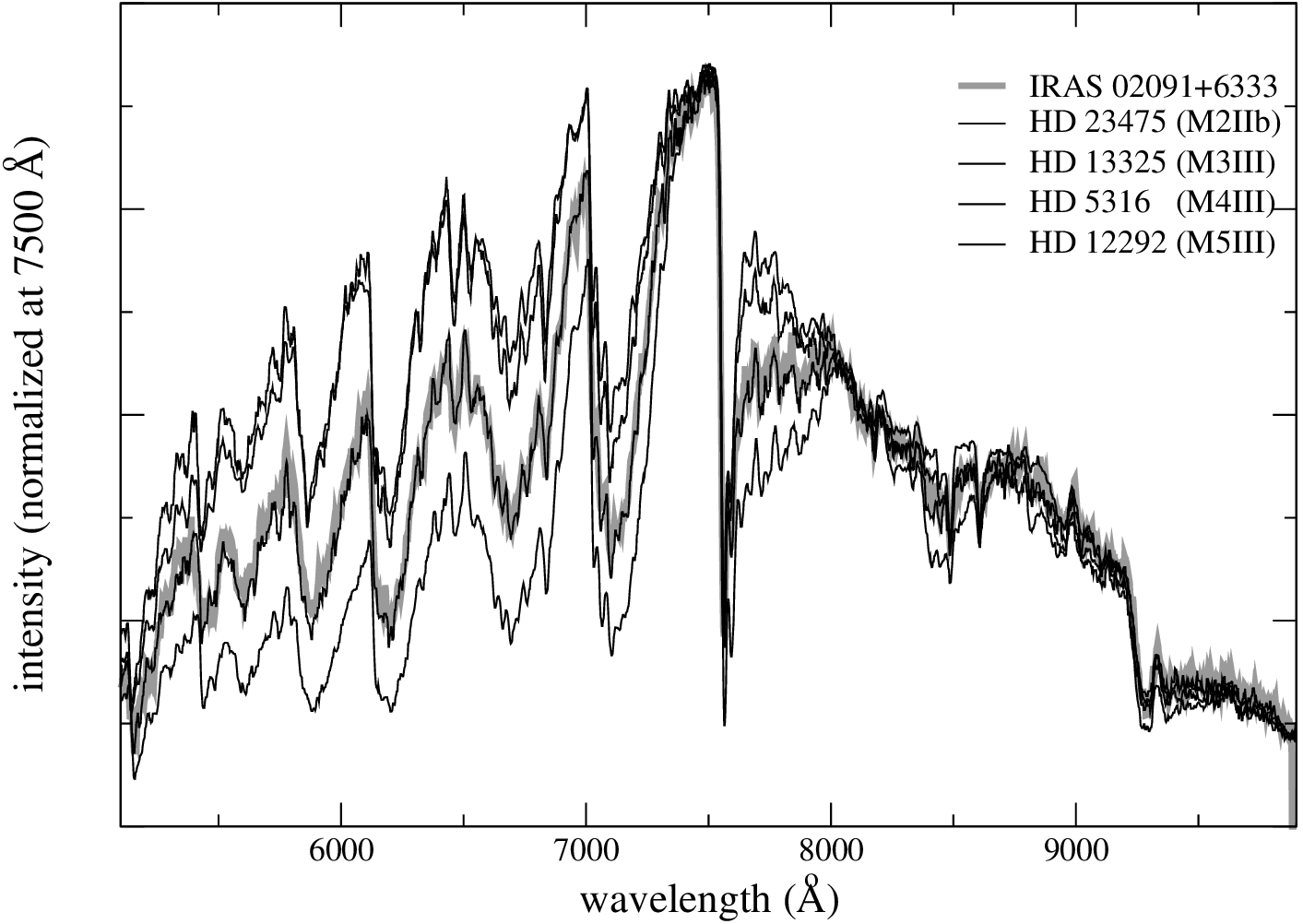}
\caption{~The composite spectrum of IRAS 02091+6333 (bold line)
using an interstellar extinction $E_{B-V} = 0\fm43$ (see text)
overlaid with (from top to bottom) the spectra of HD~23475
(M2IIb), HD~13325 (M3III), HD~5316 (M4III) and HD~12292 (M5III).
Those of HD~40239 (M3II), HD~39801 (M2I), HD~44478 (M3III) and
HD~42995 (M3III) were not plotted for clarity. The spectrum of
HD~5316 is almost completely hidden by that of the target.}
\label{fig_comparison}
\end{figure*}

The most reliable classification is the direct comparison with
standards, using the identical instrument setup. For this purpose
eight comparison stars were taken during February 2001 and January
2003 between the target exposures with the same setup. Especially
the TiO features at 7200-7400~\AA\ and 7700-8000~\AA\ are
extremely sensitive to the effective temperature of the stars in
that domain. The latter was not covered by the spectrum used by
Zijlstra \& Weinberger (\cite{ZW}). The classification using such
bands is completely independent of the correction for the
atmospheric and interstellar extinction. The results are shown in
Fig.~\ref{fig_comparison}. The target superimposes nearly exactly
the M4III star HD~5316. The overlay gives us an exact solution for
the interstellar extinction of $E_{B-V} = 0\fm43\pm0\fm01$.

We tried to use the line ratios of the IR CaII triplet of the
comparison stars and those of the target - all taken with the same
instrument setup. But not even for the comparison stars, having a
much better S/N ratio than the target, the effective resolution
(see Table~\ref{tab_obs}) allow to derive useful separations
between luminosity class III and II.

\noindent We favor a spectral type of M4III. The luminosity class
remains somewhat uncertain, but is most likely III. This is
supported also by the distance (see section \ref{phot_sec} and
discussion in Zijlstra \& Weinberger \cite{ZW}).

\subsection{Photometry and Distance}
\label{phot_sec}

\begin{table*}[ht]
\caption{Results of the Mercator photometry in the Geneva system
(Meylan \& Hauck \cite{mercator}).} \label{mercator_tab}
 {\tt
\begin{tabular}{ccccccccc}
\hline
 {\rm MJD}  &     $v$  &    $u$    &   $b$  &    $b_1$   &   $b_2$   &   $v_1$  &     $g$ & {\rm weight} \\
\hline \hline
52103.63730 & 10.445 & 15.647 & 12.027 & 13.638 & 13.011 & 11.298 & 11.281 & 4\\
52103.68906 & 10.452 & 15.540 & 12.039 & 13.646 & 12.995 & 11.291 & 11.281 & 4\\
52104.67404 & 10.450 & 15.643 & 12.017 & 13.615 & 13.007 & 11.287 & 11.286 & 3\\
52111.70459 & 10.380 & 15.416 & 11.965 & 13.585 & 12.948 & 11.232 & 11.214 & 1\\
52205.54382 & 10.427 & 15.637 & 12.030 & 13.637 & 13.035 & 11.281 & 11.267 & 2\\
52299.37930 & 10.421 & 15.427 & 12.045 & 13.700 & 13.010 & 11.280 & 11.261 & 3\\
52299.42229 & 10.419 & 15.806 & 12.060 & 13.690 & 12.997 & 11.257 & 11.253 & 3\\
52500.69875 & 10.458 & 15.606 & 12.055 & 13.659 & 13.082 & 11.304 & 11.297 & 2\\
52501.69090 & 10.453 & 15.472 & 12.047 & 13.676 & 13.067 & 11.306 & 11.286 & 2\\
52502.69127 & 10.447 & 15.714 & 12.047 & 13.680 & 13.022 & 11.303 & 11.278 & 2\\
52505.64131 & 10.434 & 15.573 & 12.027 & 13.634 & 13.034 & 11.276 & 11.254 & 3\\
52506.69780 & 10.421 & 15.552 & 12.026 & 13.650 & 13.018 & 11.279 & 11.258 & 3\\
52508.69154 & 10.430 & 15.629 & 12.040 & 13.676 & 13.017 & 11.262 & 11.264 & 2\\
\hline
\end{tabular}
}
\end{table*}

 We have searched literature and telescope archives for
photometry, but found in the optical and in the MIR-FIR only those
mentioned already in Zijlstra \& Weinberger (\cite{ZW}) -- namely
TYCHO in the optical, Midcourse Space Experiment (MSX, Egan et al.
\cite{msx}) and IRAS in the infrared. The results of our
photometry, together with results from literature are collected in
Table~\ref{photometry}. For the flux to magnitude conversion for
IRAS we used the results of Wainscoat et al. (\cite{SKY}) and for
MSX that of Cohen et al. (\cite{cohen}). The calibration of IRAS
by Hickman et al. (\cite{hickman}) was not taken into account, as
the MSX and the IRAS 12~$\mu$m fluxes gave differences of about
0\fm40 for all M type comparison stars here. For the TYCHO
measurements we applied the color equations as given in the
catalogue: $V = V_{\rm T} - 0.090 \times (B_{\rm T}-V_{\rm T})$
and $B-V = 0.850\times(B_{\rm T}-V_{\rm T})$.

During the reviewing process of this work, the 2MASS NIR data was
made public. Although we listed the values in
Table~\ref{photometry}, they were not useful, as the target was
overexposed even in the short 53~ms preexposure of the survey. We
thus didn't include these values in our interpretation.

\begin{table}[ht]
\caption{Summary of photometric observations except those listed
in Table~\ref{mercator_tab}. The TYCHO measurements were converted
to the standard system (see text).} \label{photometry}

\begin{tabular}{llccc}
\hline
  date & band  & mag. & quality  & ref.\\
 & &  & or error &  \\
 \hline \hline
 10./11.12.2002 & ~$B$  & 12\fm607 & 0\fm014 & here \\
 &  ~$V$  & 10\fm519 & 0\fm009 & here \\
 &  ~$R$  & 9\fm077 & 0\fm004 & here \\
 &  ~$I_{\rm C}$ & 7\fm309 & 0\fm019 & here \\
$\approx$ 1989 & ~$B$ & 12\fm22~ & 0\fm23~ & TYC-1 \\
 &  ~$V$ & 10\fm41~ & 0\fm05~ & TYC-1 \\
 &  ~$B$ & 12\fm46~ & 0\fm23~ & TYC-2 \\
 &  ~$V$ & 10\fm37~ & 0\fm05~ & TYC-2 \\
January 2000 & ~$B$ & 12\fm55~ & ? & [1] \\
  & ~$V$ & 10\fm5~~ & ? & [1] \\
  &  ~$R$ & ~9\fm5~~ & ? & [1] \\
  &  ~$I$ & ~7\fm4~~ & ? & [1] \\
January 2001 & ~$B$ & 11\fm7~~ & ? & [1] \\
  &  ~$V$ & 10\fm1~~ & ? & [1] \\
  &  ~$R$ & ~9\fm3~~ & ? & [1] \\
  &  ~$I$ & ~7\fm3~~ & ? & [1] \\
5.1.1999 & ~$J$ & 5\fm326$^\#$ & 0\fm03 & 2MASS \\
         & ~$H$ & 4\fm350$^\#$ & 0\fm29 & 2MASS \\
         & ~$K_{\rm s}$ & 4\fm034$^\#$ & 0\fm34 & 2MASS \\
$\approx$ 1995 &  $[$8.3$]$& ~3\fm80~ & qal 4 & MSX \\
 & $[$14.6$]$& ~2\fm9~~ & qal 1 & MSX \\
$\approx$ 1983  & $[$12$]$ & ~3\fm76~ & qal 3 & IRAS \\
 & $[$25$]$ & ~3\fm55~ & qal 3 & IRAS \\
\hline
\end{tabular}\\
\smallskip\\
\begin{tabular}{ll}
[1] & {\footnotesize  Zijlstra \& Weinberger (\cite{ZW})} \\
$^\#$ & {\footnotesize The 2MASS data is flagged in the data base
as}\\
& {\footnotesize {\sl "radial profile fitting of overexposed
source"} and}\\
& {\footnotesize {\sl "profile fit very poor"}.}\\
\end{tabular}
\end{table}

\begin{figure}[ht]
\centering
\includegraphics[width=\columnwidth]{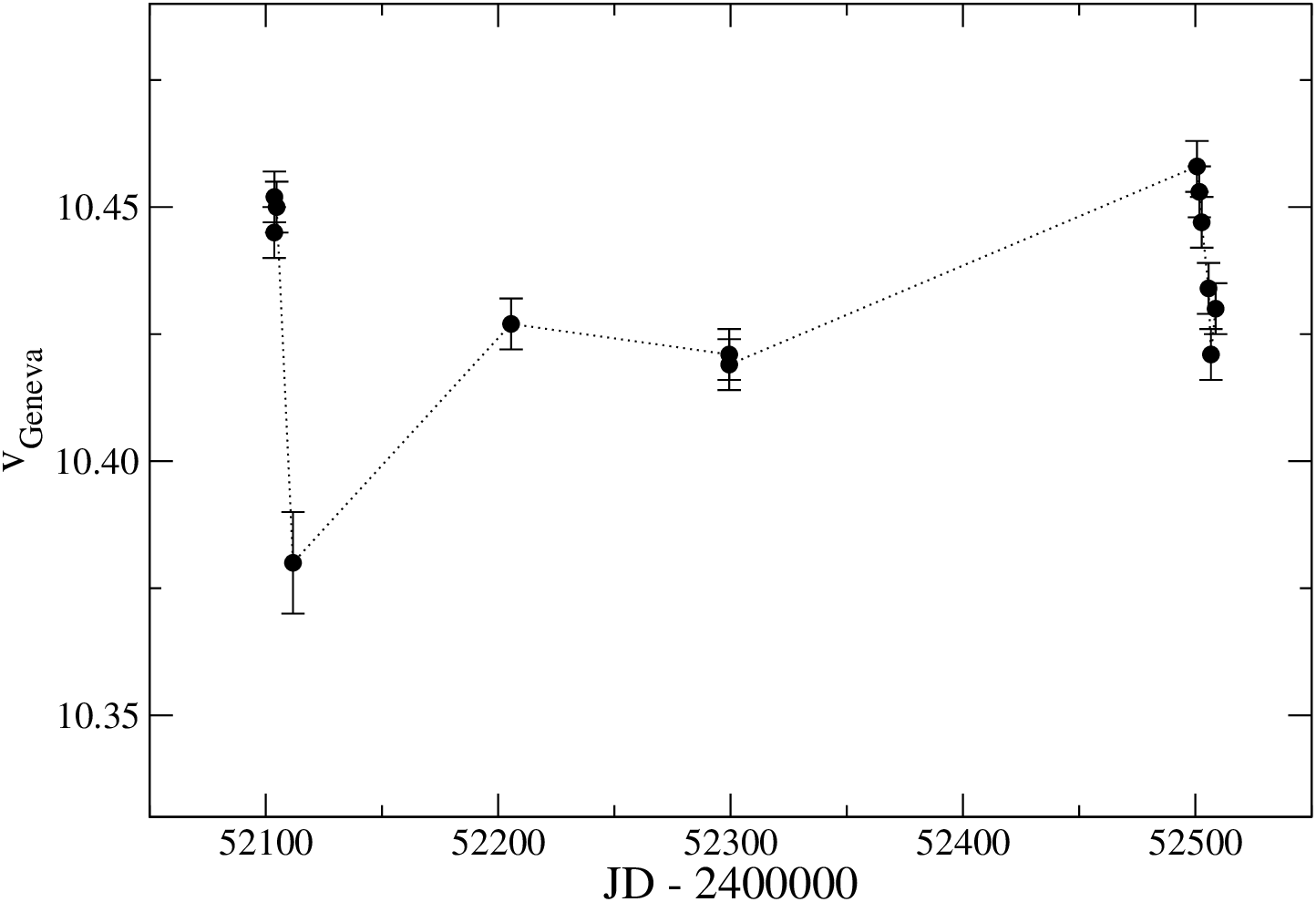}
\caption{The Mercator photometry shows a weak photometric
variation on medium to long timescales.} \label{fig_mercator}
\end{figure}

The first TYCHO catalogue (Perryman et al. \cite{tycho}) and the
reanalysis of the same data in TYCHO-2 (H{\o}g et al.
\cite{tycho2}) give completely different results for the $B_{\rm
T}$ band. The $V_{\rm T}$ band differs considerably more than the
error given in the catalogue. Thus we assume the error to be
significantly higher. For further analysis we use the TYCHO-2
values.


The Mercator monitoring (Fig.~\ref{fig_mercator}) clearly shows an
irregular variability with an amplitude below 0\fm1 in $V$.\\
On the other hand the variation between our measurements, those of
the amateur observations reported in Zijlstra \& Weinberger
(\cite{ZW}) and TYCHO-2 in the optical give an amplitude of some
tenths of a magnitude. The ($V_{{\rm T} min}$ - $V_{{\rm T} max}$)
value (15\% and 85\% of the 228 photometric transits) given in
TYCHO-2 is 1\fm71. But we assume the latter is mainly due to the
limits of the experiment. \\
The MSX and IRAS measurements together with the spectroscopic
stability lead us to the conclusion that the object seems to be
very stable in the effective temperature. That fact limits the
optical photometric amplitude to those found in the Mercator
monitoring. Such small variations seem to be usual for late type
stars around the tip of the first giant branch (Ita et al.
\cite{ita}).

\begin{figure}[ht]
\centering
\includegraphics[width=\columnwidth]{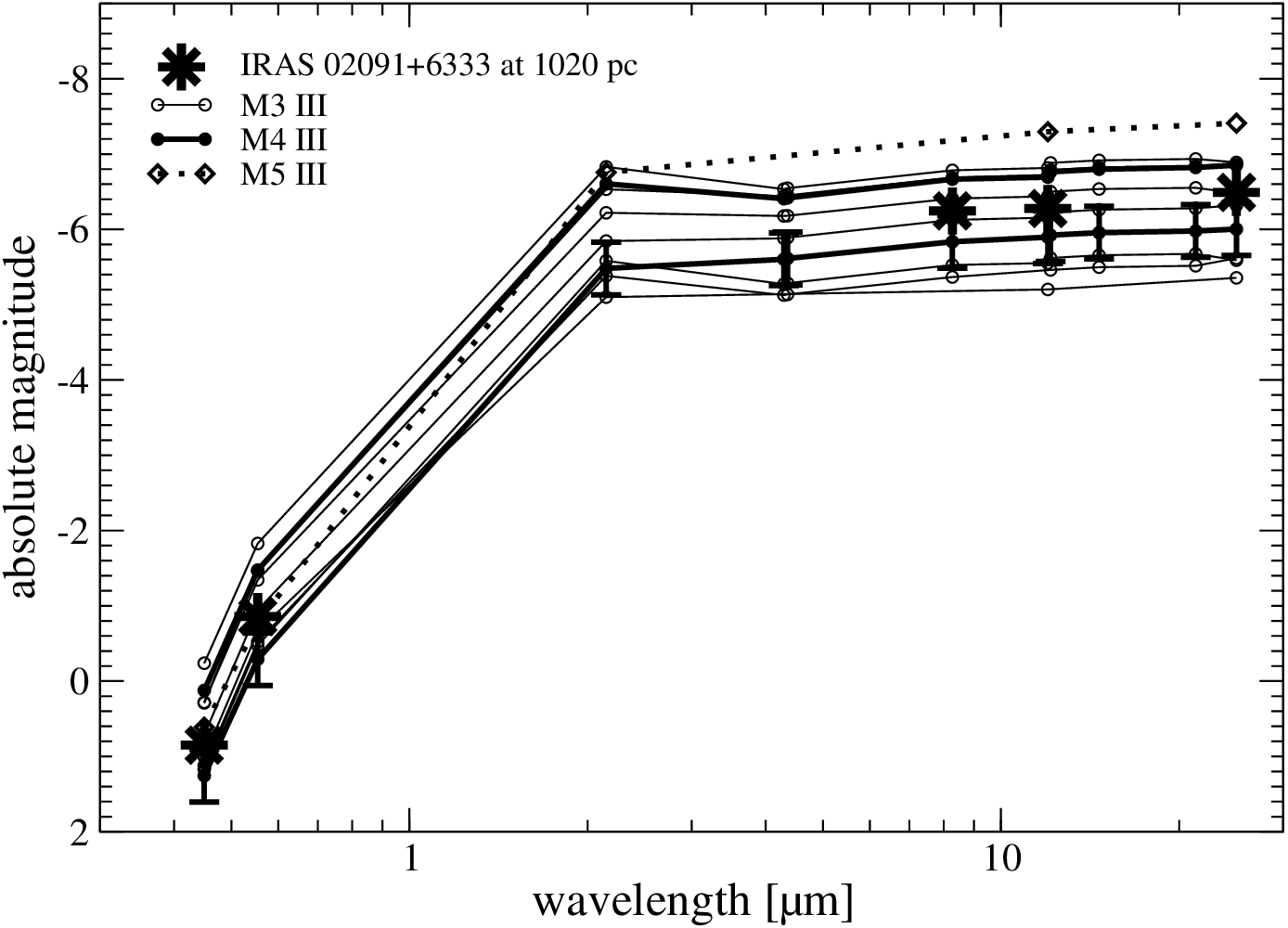}
\caption{The SEDs of the M3-5 III stars of the sample from
van~Belle et al. (\cite{belle}) and of the stars used for spectral
comparison here. The distances for calibration of the absolute
magnitude were taken from HIPPARCOS. The photometry was taken from
SIMBAD ($B$,$V$), IRC ($K$, Neugebauer \& Leighton \cite{tmss}),
MSX ($4-21\mu$m) and IRAS ($12$ and $25\mu$m). The error bars give
the tolerance added by the HIPPARCOS errors for the worst case
star. The target fits well within the two M4III stars assuming a
distance of 1020~pc. } \label{fig_IR}
\end{figure}

The colors ($V$-$R$)$_0$~=~1\fm15 and ($V$-$I_{\rm
C}$)$_0$~=~2\fm68 indicate a M4.5III MK class (Drilling \& Landolt
\cite{allen}). Using our photometry and the color equations above
we receive $V_{\rm T}-[8.3] = 5\fm56$. Cohen et al. (\cite{cohen})
are announcing 5\fm21, 5\fm58 and 6\fm02 for M3, M4 and M5
respectively, which is also consistent with the classifications
found above. The Mercator monitoring was not used for the colors,
as this system is lacking investigations, absolute calibrations
and color equations for very late type stars (Meylan \&
Hauck \cite{mercator}, Moro \& Munari \cite{asiago}).\\
 We
derived, again using TYCHO, MSX measurements and HIPPARCOS
parallaxes, the spectral energy distributions (SED) for M3-5
giants from our spectroscopic sample and from those one of
van~Belle et al. (\cite{belle}). The latter is a sample of stars
investigated in detail with interferometers for the determination
of accurate size and luminosity of late type stars. Two stars --
namely HD~13325 (M3III) and HD~5316 (M4III) -- overlap between
their and our sample. One can clearly see in Fig.~\ref{fig_IR} the
broad band mid-IR flux hardly varies between M3 and M4 stars. Only
M5 stars have a significant different SED. We thus fit our target
between the M4 stars of the sample. This provides us with (using
all bands) a distance  of 1020~pc. The HIPPARCOS errors for the
bright comparison stars and the uncertainties from the photometry
of the target results in a lower limit of 870~pc and an upper one
of 1150~pc. The luminosity derived from the spectroscopy and the
$V$ photometry give us 920~pc. According to van~Belle et al.
(\cite{belle}) for $K-[12] > 0\fm6$ there is no correlation with
the effective temperature anymore. Thus
 this band, lacking here, would deliver us no additional information.

\subsection{Astrometry}
The target was suspected to have a relatively high proper motion
in the TYCHO catalogue, although the positional error given there
is very high. We thus tried to derive a value for the proper
motion in the same way as in Kimeswenger et al. (\cite{rxs}) by
using the six sky survey plates of the region. But as the target
is extremely overexposed and lies near the corner of the POSS-E
(1952 and 1993) and of the Quick-VN (1983) plates, we cannot
derive a proper solution. The accuracy is about the same as in the
the TYCHO experiment. Kislyuk et al. (\cite{fonac}) give similar
numbers in their online version, using their plate from 1983 and
the AC2000 catalogue position from 1905 (Urban et al.
\cite{ac2000}). But forwarding the errors in both catalogues to
the resulting proper motion we again end up with large error bars
(12$\pm7$~mas~yr$^{-1}$). The direction of the motion vector is
co-lined with the TYCHO result, suggesting a genuine movement
which leads to a motion of about 60~km~s$^{-1}$ relative to the
surrounding stellar field after subtracting the galactic rotation
for a distance of about 1~kpc (only about 1.5~mas~yr$^{-1}$). This
is, although still possible, very high for galactic disk stars.
The vector is almost opposite to the galactic rotation parallel to
the galactic plane. Further investigations about the population
membership are needed.

\section{Conclusions}
\begin{table}[ht]
\caption{Summary of the results for IRAS 02091+6333}
\begin{tabular}{ll}
spectral type & M4$_{\pm 0.3}$ III \\
T$_{\rm eff}$ & 3350$_{\rm \pm 50}$~K \\
type of photometric variability & irr. or SR \\
$<V>_{\rm Johnson}$ & 10\fm52$_{\pm 0.03}$ \\
interstellar $E_{B-V}$ & $0\fm43_{\pm0.01}$ \\
visual amplitude & $\le$ 0\fm1 \\
M$_V$ & $-$0\fm5 \\
M$_{\rm bol}$ & $-$2\fm65 \\
distance      & $1020^{+130}_{-150}\,\,$pc \\
luminosity  & $920_{\pm 150}\,\,$L$_\odot$ \\
\end{tabular}
\end{table}

 The precise spectroscopic classification of the central star
IRAS~02091+6333 of the unique giant dust shell reported in
Zijlstra \& Weinberger (\cite{ZW}) allowed us to derive a more
accurate distance of the target. This leads us to a somewhat
higher mass estimate of 5.2~M$_\odot$ and a size of
1.8~$\times$~10$^{17}$~m (5.2~pc) and thus even enlarges the
discrepancy in lifetime as discussed in section 3.3 of Zijlstra \&
Weinberger (\cite{ZW}) and therefore clearly excludes the scenario
of the mass originating mainly from the star itself. This result
is strengthened by the fact of the lower luminosity we derived
(40\% below the one estimated there). The star lies in the
evolutionary tracks (Girardi et al. \cite{tracks}) nearby the red
giant branch tip and rather far from the TP-AGB. The small
irregular photometric variability fits well to this assumption
(Ita et al. \cite{ita}). So the model of the swept up ISM is
strongly supported by our measurements. The fast movement,
although uncertain, may shorten the timescales for the bubble.
This may suggest a wind having a higher momentum. One can only
speculate on the origin of this additional source of energy. There
is no signature in the spectroscopy for a hot companion causing a
symbiotic Mira. Also Schmeja \& Kimeswenger (\cite{schmeja}) have
shown that those objects have clear signatures in their NIR and
MIR bands.

IRAS~02091+6333 has a distance of $D = 1020^{+130}_{-150}\,\,$pc,
a luminosity of $L = 920\pm150\,\,$L$_\odot$ and, according to
van~Belle et al. (\cite{belle}), an effective temperature of
$T_{\rm eff} = 3350\pm50\,\,$K. Both the spectroscopy as well as
the photometric consistency of the colors with the MK type from
visual to IR bands suggest there is currently no dominant
circumstellar shell material. Girardi et al. (\cite{girardi})
discussed inhomogeneities in the tracks and in the evolution due
to different He-core burning phases. It affects slightly the
luminosity (and thus the distance) but hardly the photometric
colors (Girardi et al. \cite{tracks}). This adds some
uncertainties to the values derived for individual sources.

\begin{acknowledgements}
We thank the colleagues of the institute for the unlimited access
to the facilities of the new university observatory in Innsbruck.
The staff of the 'Instituut voor Sterrenkunde', KULeuven is
acknowledged for the photometric monitoring at the Mercator
facilities and the staff of the Geneva observatory is acknowledged
for the photometric reduction

\end{acknowledgements}

\end{document}